\documentclass[traditabstract]{aa} % for the abstract without structuration 
\usepackage{graphicx,rotating,amssymb}
\usepackage{booktabs}
\usepackage{subfig}
\usepackage{float}
\usepackage{natbib}
\usepackage{multirow}
\usepackage[english]{babel}
\usepackage{txfonts}
\begin{document}
   \title{The physics and the structure of the quasar-driven outflow in Mrk~231}

   \author{C. Cicone
          \inst{1, 2} 
          \and
          C. Feruglio \inst{3}
         \and
         R. Maiolino  \inst{1, 2}
	\and
        F. Fiore  \inst{1}
	\and
	E. Piconcelli \inst{1,4}
	\and
	N. Menci \inst{1}
	\and
        H. Aussel \inst{5}
	\and
         E. Sturm \inst{6}
	}

      \institute{Cavendish Laboratory, University of Cambridge 19 J. J. Thomson Avenue, Cambridge CB3 0HE, UK\\
           \email{c.cicone@mrao.cam.ac.uk}
	  \and INAF - Osservatorio Astronomico di Roma (OAR), via Frascati 33, 00040 Monteporzio Catone, Italy\\
	  \and Institut de Radio Astronomie Millim\'{e}trique (IRAM), 300 rue de la Piscine, St. Martin d'H\`{e}res, France\\
	  \and XMM-Newton Science Operations Centre, ESA, P.O. Box 78, 28691 Villanueva de la Cañada, Madrid, Spain \\
            \and Laboratoire AIM, DSM/Irfu/Service d'Astrophysique, CEA Saclay, 91191 Gif-sur-Yvette, France \\
	    \and Max Planck Institute f\"{u}r Extraterrestrische Physik (MPE), Postfach 1312, D85741 Garching, Germany
	      }
	
   \date{Accepted: 22 May 2012}

 \abstract{Massive AGN-driven outflows are invoked by AGN-galaxy co-evolutionary models to
suppress both star formation and black hole accretion. Massive molecular outflows
have been discovered in some AGN hosts. However, the physical
properties and structures of these AGN-driven molecular outflows are still poorly constrained. Here
we present new IRAM PdBI observations of Mrk~231, the closest quasar known, targeting
both the CO(1-0) and CO(2-1) transitions. We detect broad wings in
both transitions, which trace a massive molecular outflow moving with velocities of up to 800
km/s. The wings are spatially resolved at high
significance levels (5--11$\sigma$),
indicating that the molecular outflow extends to the kpc scale. 
The CO(2-1)/CO(1-0) ratio of the red broad wings is 
consistent with the ratio observed in the narrow core, while
the blue broad wing is less excited
than the core. The latter result suggests that quasar-driven outflow models invoking shocks
(which would predict higher gas excitation) are inappropriate for describing the bulk of the outflow
in Mrk~231. However, we note that within the central 700 pc the CO(2-1)/CO(1-0) ratio of the red
wing is slightly, but significantly,
higher than in the line core, suggesting that shocks may play a role in the
central region.
%at least for
%part of the outflow in Mrk231.
We also find that the average size of the
outflow anticorrelates with the critical density of the transition used as a wind
tracer. This indicates that, although diffuse and dense clumps coexist in the
outflowing gas, dense outflowing clouds have shorter lifetimes and that they
evaporate into the diffuse
component along the outflow or, more simply, that diffuse clouds are more efficiently
accelerated to larger distances by radiation pressure.}
 
 
   \maketitle

%
%________________________________________________________________

\section{Introduction}

The ubiquitous discovery of ``relic'' supermassive black holes (SMBH) in local bulges and 
of the tight correlations between their masses and bulge properties 
(e.g. \citealt{Ferrarese+Ford05}; \citealt{Marconi+Hunt03}; \citealt{Gultekin+09}; and references therein)
indicates that nuclear activity likely plays an important role in galaxy evolution.
Considerable attention has been drawn to powerful AGN-driven
outflows as they may provide a very efficient mechanism enabling the nuclear
 activity occurring on sub-parsec scales
to influence the surrounding host galaxy. 
The effect of these AGN winds on the interstellar matter (ISM) in the
galaxy is often referred to as
``AGN feedback''. More specifically,  AGN feedback is invoked by most current galaxy evolutionary models
to explain the shortage of very massive galaxies in the local Universe,
their red colors, and low gas content, as well as the $\rm M_{BH}-\sigma$ relation 
(e.g. \citealt{Silk+Rees98}, \citealt{Fabian99},
\citealt{King+Pounds03}; \citealt{Fabian+06}
\citealt{Baldry2+04}; \citealt{Granato+04}; \citealt{DiMatteo+05}; \citealt{Elvis06};
\citealt{Lapi+05};  \citealt{King05}; \citealt{Menci+06, Menci+08}; \citealt{Croton+06}; 
\citealt{Narayanan+08};  \citealt{Hopkins+Elvis10}).

There are two possible ``flavors'' of AGN feedback: feedback associated
with the ``radio-mode'' \citep{Croton+06}
and feedback associated with the ``quasar-mode''.
In this work, we
focus only on the second one, which is the impulsive feedback exerted
by AGNs during their bright active phases.
One of the most common scenarios
in theoretical models of quasar-mode feedback is that nuclear
fast winds, generated by the AGN radiation pressure,
deposit the energy produced by the nuclear activity into the ISM, 
compressing the ISM into a shock wave (``blast wave'')
that eventually blows the cold gas reservoir, available for both
star formation and SMBH accretion, out of the host galaxy
(\citealt{Lapi+05}; \citealt{Menci+06, Menci+08};  \citealt{King10},
\citealt{Zubovas+12}). Models generally assume that these outflows also contain a significantly
large molecular component (\citealt{Zubovas+12}, \citealt{Narayanan+08}).
If AGN-driven outflows can really expel large quantities of molecular gas
(which constitutes the bulk of the ISM in the central regions and is the medium out
of which stars form), then this would be an extremely effective mechanism for quenching
star formation in massive galaxies.

Major observational breakthroughs in this field have been achieved  
by revealing massive and energetic quasar-driven outflows in several galaxies,
both locally and at high redshift \citep{Fischer+10,Feruglio+10,Rupke+11,Aalto+11,
Sturm+11,Alatalo+11,Greene+11,
Nesvadba+10,Nesvadba+11,Alexander+10,Farrah+12,Cano-Diaz+12,Maiolino+12} (see also
the review by \citealt{Fabian12}). Within this
context, Mrk~231, the closest quasar known, is the object in which a massive quasar-driven
outflow was first discovered and that has been studied in the greatest detail
(\citealt{Fischer+10}; \citealt{Feruglio+10};  \citealt{Rupke+11}, \citealt{Aalto+11}). 
Mrk~231 provides a unique, nearby laboratory for investigating AGN feedback, showing all of the 
typical features expected for a quasar transiting from the obscured, merger-driven and IR-luminous 
accretion phase, accompanied by strong circumnuclear star formation, to the unobscured, 
standard  phase of a 'blue' quasar (\citealt{Lipari+09} and references therein), 
by expelling obscuring gas and dust. 
It is indeed  the most luminous ultra-luminous infrared galaxy (ULIRG) in the local Universe
\citep{Sanders+03},
with infrared (IR) luminosity L$_{IR}$(8-1000$\mu$m) = 1.33$\times10^{46}$ erg s$^{-1}$,
%$L_{IR}(8-1000\,\mu m)= 3.5 \times 10^{12}\,L_{\odot}$, 
and also hosts
 a very powerful  ($L_{bol,\,AGN} = 1.1 \times 10^{46}$ erg s$^{-1}$; e.g., \citealt{Rupke+11}), 
low-ionization broad absorption line (LoBAL) quasar  (i.e. where the blue-shifted UV lines
span velocities up to $\sim 8000$ km s$^{-1}$). 
\citet{Boroson+92} argued that LoBAL quasars are young, heavily enshrouded AGNs, where the cocoon of gas 
and dust has a large covering factor,  
as supported by the results of \citet{Braito+04}, who found that Mrk 231 is 
a very obscured X-ray source with a column density of $N_H$ $\approx$ 10$^{24}$ cm$^{-2}$.
Mrk 231 also displays a distorted optical morphology, which implies that its host galaxy 
is in an advanced stage of a merging process \citep{Lipari+09}. 
The circumnuclear starburst in this source is very young ($\lesssim 120$ Myr), with an estimated
star formation rate of SFR  $\approx 200$ M$_{\odot}$ yr$^{-1}$
\citep{Taylor+99,Davies+04}. 

A Herschel PACS spectrum of Mrk 231 has revealed a remarkably massive
molecular outflow traced by prominent P-Cygni OH profiles at 65, 79, and 119 $\mu$m with velocity shifts of
$\sim$ 1000 km s$^{-1}$ \citep{Fischer+10}. 
In \citet{Feruglio+10}, we reported the detection of the molecular outflow based on IRAM PdBI observations, 
which revealed the broad (FWZI$\sim$1500 km s$^{-1}$) wings
of the CO(1-0) transition, that are marginally resolved with an extension of about $\sim$ 1 kpc, 
yielding an estimated mass-outflow rate of $\sim$700 M$_{\odot}$ yr$^{-1}$, i.e., far higher than the
ongoing SFR in Mrk 231. On the basis of kinetic power arguments, we also suggested that the molecular outflow
is primarily driven by the quasar radiation pressure.

These findings were later confirmed by the analysis of the broad wings
of the  H$\alpha$ and Na I D absorption line via IFU optical spectroscopy, 
which spatially resolved the outflow on the same
scales as traced by the CO wings \citep{Rupke+11}. 
Furthermore, \citet{Aalto+11} also detected broad 
wings in the HCN(1-0), HCO$^{+}$(1-0), and HNC(1-0) emission lines of Mrk 231, which are tracers of the
high-density molecular gas, and confirmed the extent of the HCN(1-0) broad emission out to
0.7 kpc. They suggested that the molecular outflow in Mrk 231 is clumpy and dominated by 
the dense phase, and has an enhanced HCN abundance, which may indicate either that it contains a shocked medium in the outflow 
and/or a chemistry influenced by the AGN.
Although these characteristics ensure that Mrk~231 is an exception in  the
local Universe, they are expected to be quite common at high redshift, where
 Mrk~231-like objects 
are thought to play a major role in the formation of present-day, red massive quiescent 
elliptical galaxies (\citealt{Hopkins+08}; \citealt{Cattaneo+09}).
We note, in particular, that massive molecular outflows have been revealed in additional local ULIRGs
through the detection of prominent OH P-Cygni profiles in their far-IR Herschel 
spectra \citep{Sturm+11}.
%detected 
%in other active ULIRGs \citep{Sturm+11} by revealing OH P-Cygni profiles, i.e.,
%with blueshifted absorption and redshifted emission.

In this paper we present the analysis of a new set of  IRAM PdBI observations of the
CO(1-0) and CO(2-1) transitions in Mrk 231.
The aim of this study is
 to improve the previous results achieved by
\citet{Feruglio+10}, by spatially resolving the molecular outflow with
high significance (5--11$\sigma$).
Moreover, we combine the CO(1-0) and CO(2-1) emission line data to
investigate the physical conditions of the molecular
outflowing gas in this unique source.

A $H_0$=70.4 km s$^{-1}$ Mpc$^{-1}$, $\Omega_M=0.27$, $\Omega_{\Lambda}=0.73$ 
cosmology is adopted throughout this work.

\section{Observations}

Mrk~231 was observed in CO(1-0) and CO(2-1) with the IRAM millimeter-wave interferometer on
Plateau de Bure (IRAM PdBI) between June 2009 and November 2010. The CO(2-1) 
observations presented allow us to investigate the broad wings in Mrk231 for the first time in
this transition. The new CO(1-0) instead allows us to improve on the previous Feruglio et al. (2010) observation
by significantly increasing the signal on long baselines and, therefore, allowing us to investigate
in more detail the extension of the outflow in this transition.
Table \ref{table:data} provides a technical description of these observations:
the CO transition and the corresponding redshifted frequency (z=0.04217), 
the dates of the observations, the array configurations and the number
of antennas used, the on-source integration times, the beam sizes, and the relevant references.

The data were calibrated by using the CLIC software of the GILDAS package \footnote{http://www.iram.fr/IRAMFR/GILDAS}.
The main flux calibrator at 110.697 GHz was MWC 349, whose flux at this frequency is 1.27 Jy; the
principal flux calibrators at 221.210 GHz were instead 0923+392 and 3C84, whose fluxes at this frequency
are 3.11 Jy and 7.71 Jy, respectively.
The absolute flux calibration obtained is typically precise to better than $10\%$ at 110.697 GHz and $\sim20\%$ at 221.210 GHz
\citep{PdBI_cookbook}.
After calibration, old (2009) and new (2010) data of the CO(1-0) emission line were merged together.

Both the narrow-band and the wide-band (WideX) correlators offered by the PdBI were exploited in our measurements.
The WideX provides a spectral resolution of 1.95 MHz over its full bandwidth of 3.6 GHz and is
available in parallel to the narrow-band correlator. The narrow-band correlator, whose maximum signal
bandwidth is 2 GHz, was configured to have a spectral resolution of 2.5 MHz.
We rebinned the uv-tables obtained with the narrow-band correlator to give frequency channels of 12.5 MHz for 
CO(1-0) and 25 MHz for CO(2-1).
The wide-band data, thanks to
the larger number of line-free channels available, were exploited to produce uv-tables of the
continuum emission (sampled from channels with velocities $\rm |v|>1200~$km s$^{-1}$), which was
then subtracted from both the CO(1-0) and CO(2-1) narrow-band
and Widex data.
 
Data imaging, cleaning, and analysis were performed using the MAPPING
software, which is included in the
GILDAS package.
%%We made maps of the broad emission over the velocity ranges, eliminating the central km s^{-1}.
The field of view of the maps is 2.7$\times$2.7 arcmin$^2$ at 110.697 GHz and 1.3$\times$1.3 arcmin$^2$ at 221.210 GHz.
We extracted the CO(1-0) and CO(2-1) spectra from the cleaned and continuum-subtracted
narrow-band data cubes with an aperture of a diameter of 6 arcseconds.
%using two different
%circular apertures with diameters of 6 and 15 arcseconds.
The noise levels per channel of the CO(1-0) and CO(2-1) spectra are 0.4 mJy and 1.2 mJy, respectively.
We verified that the noise distribution is Gaussian by analyzing the 
distribution histogram of the fluxes. 

%Table1: IRAM PdBI data infos
\begin{table}
\caption{Description of the IRAM PdBI CO observations} % title of Table
\label{table:data}      % is used to refer this table in the text
\centering        
\scriptsize                % used for centering table
\begin{tabular}{lccccc}       % centered columns (4 columns)
\hline\midrule                 % inserts double horizontal lines
 Line 		  & Date(s)          & Conf.            & On source   & Beam 		& Ref. \\ %table headings
 (freq.)           &                  & (no. of ant.)    &   time      & (arcsec)	&  \\
\midrule
CO(1-0)           & June-Nov.  	& C+D      & 	$20$ hrs	&  $3.2 \times 2.8 $    & Feruglio  \\
(110.607 GHz)     &  09      	& (5 ant.) &			&			& et al. 2010		\\
		 &  Oct. 10    	 & C          & 	$7.4$ hrs & 			& This work   \\
       	         &   	           	 & (5/6 ant.) &                &			&		 \\
\midrule
CO(2-1)		 & Sep.-Nov.       & C+D        & 	$4.2$ hrs & $1.6 \times 1.3$	&    This work   \\
(221.210 GHz)     &    10         & (5/6 ant.) &          	  &			&    \\
\midrule   %inserts single line 
\end{tabular}
\end{table}

\section{Results}

%Figure1: three line profiles with gaussian fits \label{fig:Fig1}
\begin{figure}[!]
   \centering
  \includegraphics[width=.95\columnwidth]{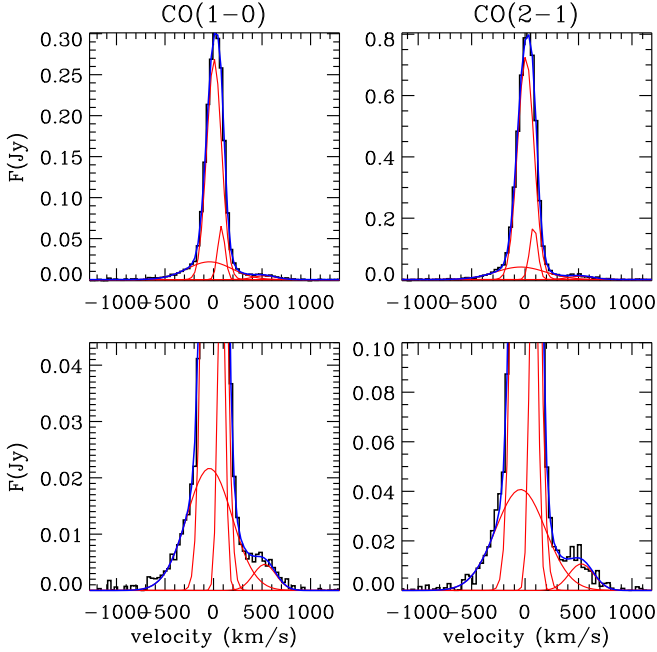}
   \caption{Continuum-subtracted 
IRAM PdBI spectra of the CO(1-0) and CO(2-1) emission lines.
These spectra are extracted from circular apertures with diameters of 6 arcseconds.
 \emph{Top panels}: full flux scale. \emph{Bottom panels}: expanded flux scale to highlight the broad components.
The two emission lines were simultaneously fitted with four Gaussian functions (red profiles, see text)
tied to have the same centers and widths for the two transitions, to reproduce the three main components
of each line (i.e. core, blue, and red broad components).
The blue contours represent the results of the fit. 
%The \textit{FWHM} of the core is $(179.7 \pm 2.6)\,km\,s^{-1}$, while
%the \textit{FWHM} of the bluer and the redder broad components are $(548\pm 35)\,km\,s^{-1}$ and $(276\pm 71)\,km\,s^{-1}$, 
%respectively.
} \label{fig:Fig1} \end{figure}

\begin{table}[tbp]
\caption{Integrated fluxes}
\label{table:3fluxes}
\centering
\scriptsize 
\begin{tabular}{lcccc}

\hline\midrule
\multicolumn{5}{c}{Fluxes from uvfits in velocity intervals} \\
\midrule
  		& \multicolumn{2}{c}{Velocity range}            & F[CO(1-0)]         	&  F[CO(2-1)]   \\
		& \multicolumn{2}{c}{(km s$^{-1}$)}		& (Jy km s$^{-1}$)	& (Jy km s$^{-1}$) \\
\midrule
 Narrow core   		& \multicolumn{2}{c}{(-300 $\div$ 300)}	& ($64.39 \pm 0.13$)     & $(194.00 \pm 0.50)$ \\
 Blue wing     	& \multicolumn{2}{c}{(-1000 $\div$ -400)}	& ($1.84 \pm 0.16$)	& ($3.90 \pm 0.38$) \\
 Red wing     	& \multicolumn{2}{c}{(400 $\div$ 1000)}		& ($2.32 \pm 0.17$)	& ($7.08 \pm 0.84$) \\
\midrule
 Total flux     & \multicolumn{2}{c}{(-1000 $\div$ 1000)}	& ($75.98 \pm 0.22$)	& ($241.32 \pm 0.66$) \\
\midrule
\multicolumn{5}{c}{Fluxes of Gaussian components} \\
\midrule
Component 	&	Velocity	& FWHM 			& F[CO(1-0)]        	&  F[CO(2-1)]   \\
		&	(km s$^{-1}$)	& (km s$^{-1}$)		& (Jy km s$^{-1}$)	& (Jy km s$^{-1}$) \\
\midrule
\multirow{2}{*}{Narrow core$^{\dag}$}  & ($6.5\pm 2.0$)	& ($179.7 \pm 2.6$) & \multirow{2}{*}{($54.6 \pm 1.1$)} & \multirow{2}{*}{($ 173.2\pm 3.3$)} \\
		& ($80.9 \pm 1.3$)	& ($86.3 \pm 5.3$)	&     &  \\
Bluer comp.	&  ($-42 \pm 16$)	& ($548\pm 35$)		& ($24.9 \pm 2.5$)	& ($52.7 \pm 6.6$) \\
Redder comp.	&  ($527\pm 30$)	& ($276\pm 71$)		& ($1.96 \pm 0.43$)	& ($6.1 \pm 1.4$) \\
\midrule
\multicolumn{3}{c}{Total flux}   				& ($75.7 \pm 1.9$)	& ($240.7 \pm 5.6)$ \\
\midrule
 \end{tabular}

\begin{flushleft}
\small
\textbf{Notes:} 
Note that a flux calibration
systematic error equal to the 10$\%$ 
of the flux value should be added
in quadrature to the statistical error reported with each flux
measurement in this table.\\
$^\dag$ The FWHM and the velocity of both the Gaussians fitting the narrow core are
reported.
\end{flushleft}
 \end{table}

Figure \ref{fig:Fig1} shows the IRAM PdBI spectra of the CO(1-0) and CO(2-1) emission lines extracted
from an aperture of 6 arcsec.
The broad component, discovered in the (1-0) transition by \citet{Feruglio+10}
is also clearly present in the (2-1) transition.
%The wings of the CO(1-0) and CO(2-1) transitions show a striking similarity. 
The spectra in Figure \ref{fig:Fig1} are fitted by using two narrow Gaussians 
to fit the core of the CO lines and two broad Gaussians to fit the wings. 
In the fitting, the centers and widths of the components are constrained to be identical for the 
two CO transitions, but their relative strengths are allowed to vary. 
The need to use two narrow Gaussians to fit the core probably reflects that
the gas distribution in the central rotating disk cannot be described
through a simple, single Gaussian.

The fluxes extracted from the cleaned cubes may be affected by various
uncertainties associated with the cleaning process and aperture effects.
Therefore, to estimate rigorously the fluxes of the components, we
used the fluxes inferred directly from the visibility data, which are 
unaffected by any cleaning problem.
More specifically, we measured the flux inferred from the uv amplitudes at the 25m baseline,
for which the beam width is 14 arcsec at 230 GHz, and is even larger at lower frequencies:
this therefore ensures that all of the flux in the wings (which
are much less extended, as discussed below) is accounted for.
To obtain this information, we extracted, for both the CO transitions,
the amplitude uv diagrams of each spectral line component separately, by averaging
the frequency channels corresponding to 
$-300\lesssim$v(km s$^{-1}$)$\lesssim 300$ (line core),
$-1000\lesssim$v(km s$^{-1}$)$\lesssim -400$ (blue wing), and
$400\lesssim$v(km s$^{-1}$)$\lesssim 1000$ (red wing). 
All three of the components are continuum-subtracted.
In addition, for the visibilities of the narrow core,
we also subtracted the contribution of the ``near'' wings
estimated within $-600\lesssim$v(km s$^{-1}$)$\lesssim -400$ and
$400\lesssim$v(km s$^{-1}$)$\lesssim 600$.
In Table \ref{table:3fluxes}, we report the fluxes of the CO lines, integrated within these
velocity intervals in the uv-amplitude data, along with
the total line fluxes estimated within $-1000\lesssim$v(km s$^{-1}$)$\lesssim 1000$,
by applying the same method\footnote{ Note that the errors reported in Table 2 are statistical errors only. 
These are the ones required to infer the statistical significance and compare the relative
intensity of the various components within the same transition. However, when comparing
with other observations or among different transitions, a systematic error of 10\% in
the absolute calibration should be added in quadrature.}.
The wings are detected at a confidence level ranging from 8$\sigma$ to 14$\sigma$, depending
on the transition.

The full width at half maximums (FWHMs), velocities, and fluxes of the Gaussian components 
resulting from the fit shown
in Figure \ref{fig:Fig1}, are also reported in Table \ref{table:3fluxes}.
The fluxes of the Gaussian components were corrected for aperture and cleaning 
effects by rescaling their fluxes
to the values obtained from the uv-amplitudes in the following way: we integrated
the spectra shown in Figure \ref{fig:Fig1} in the same velocity intervals
as those used for the uv-amplitudes of the narrow core and the broad wings, and then
estimated the three different scaling factors from the ratios of
the fluxes resulting from the uv-amplitudes to the fluxes obtained by integrating
the spectra of the individual Gaussian components. We note that the scaling factors obtained for the broad wings
are slightly higher than those obtained for the narrow core, since the spectra
extracted from an aperture with a diameter of only 6 arcsec miss
part of the broad and faint components of the two CO emission lines.

%Figure 2: maps of CO(1-0) and CO(2-1) wings
\begin{figure}[h!]
\centering
{\includegraphics[width=.42\columnwidth, angle=270]{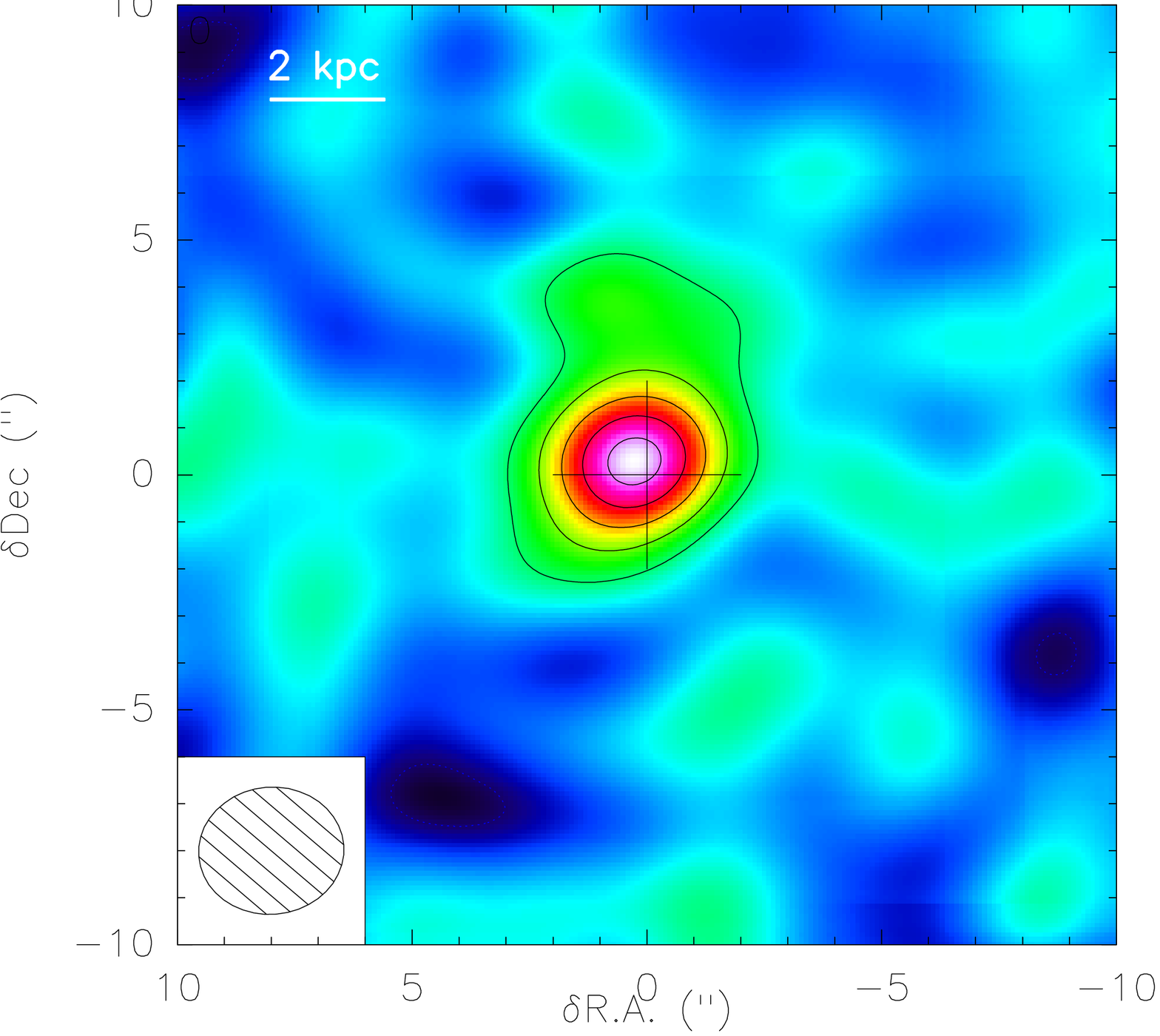}} \quad
{\includegraphics[width=.42\columnwidth, angle=270]{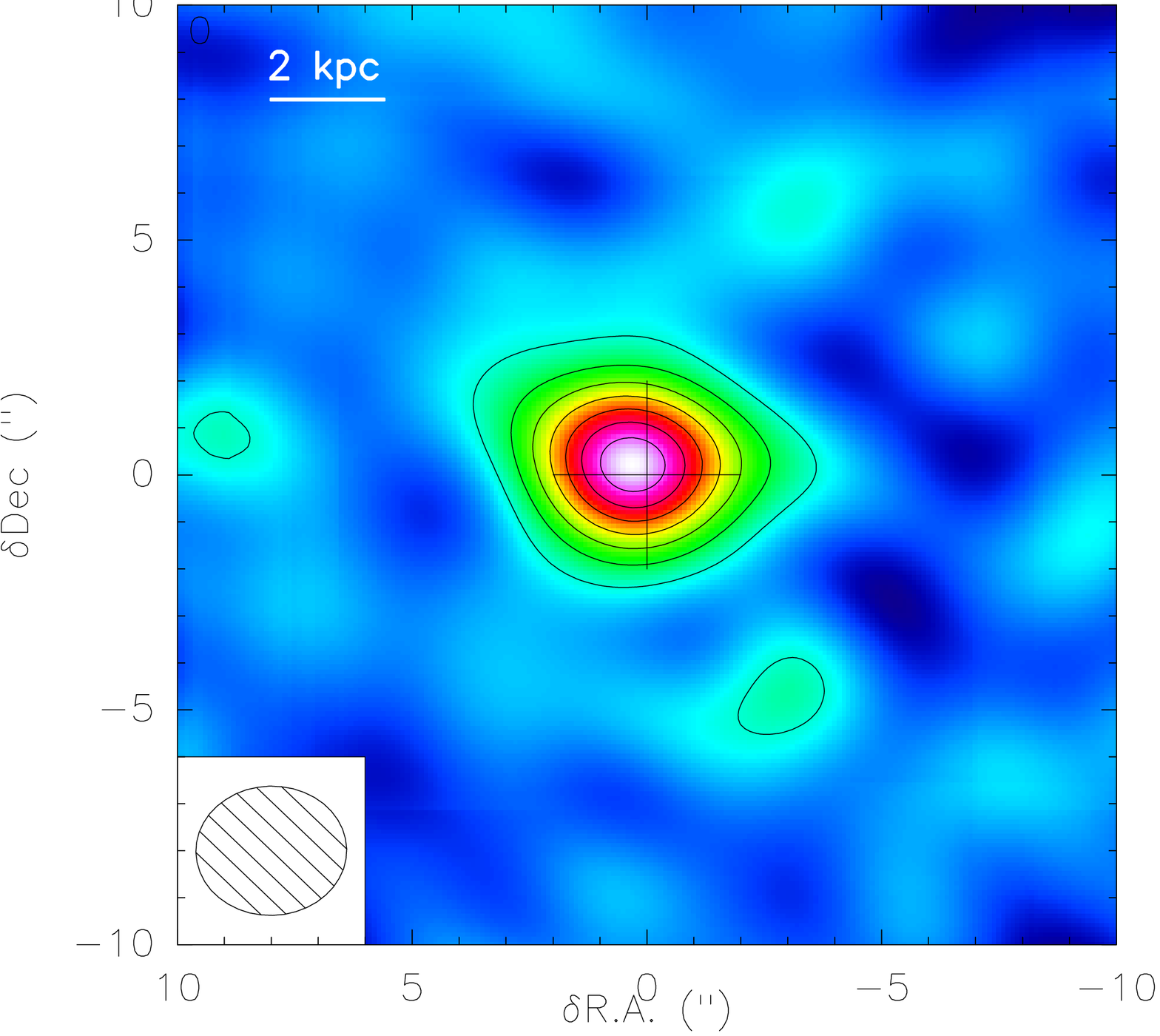}} 
\caption{IRAM PdBI maps of the CO(1-0) blue (\emph{left panel}) and red (\emph{right panel}) broad wings.
The size of the map is $20\times 20$ arcseconds. Contours correspond to 3$\sigma$ (1$\sigma$ = 0.14 mJy beam$^{-1}$).
The synthesized beam size is shown at the bottom of the maps.
%($3.2\times 2.7$ arcseconds).
The cross indicates the peak of the radio (VLBI) emission. Note that the peaks 
of the wing maps appear to be slightly offset from the peak of the radio VLBI 
but are well within the beam. } \label{fig:wing10map} \end{figure}
\begin{figure}[h!]
\centering
{\includegraphics[width=.42\columnwidth, angle=270]{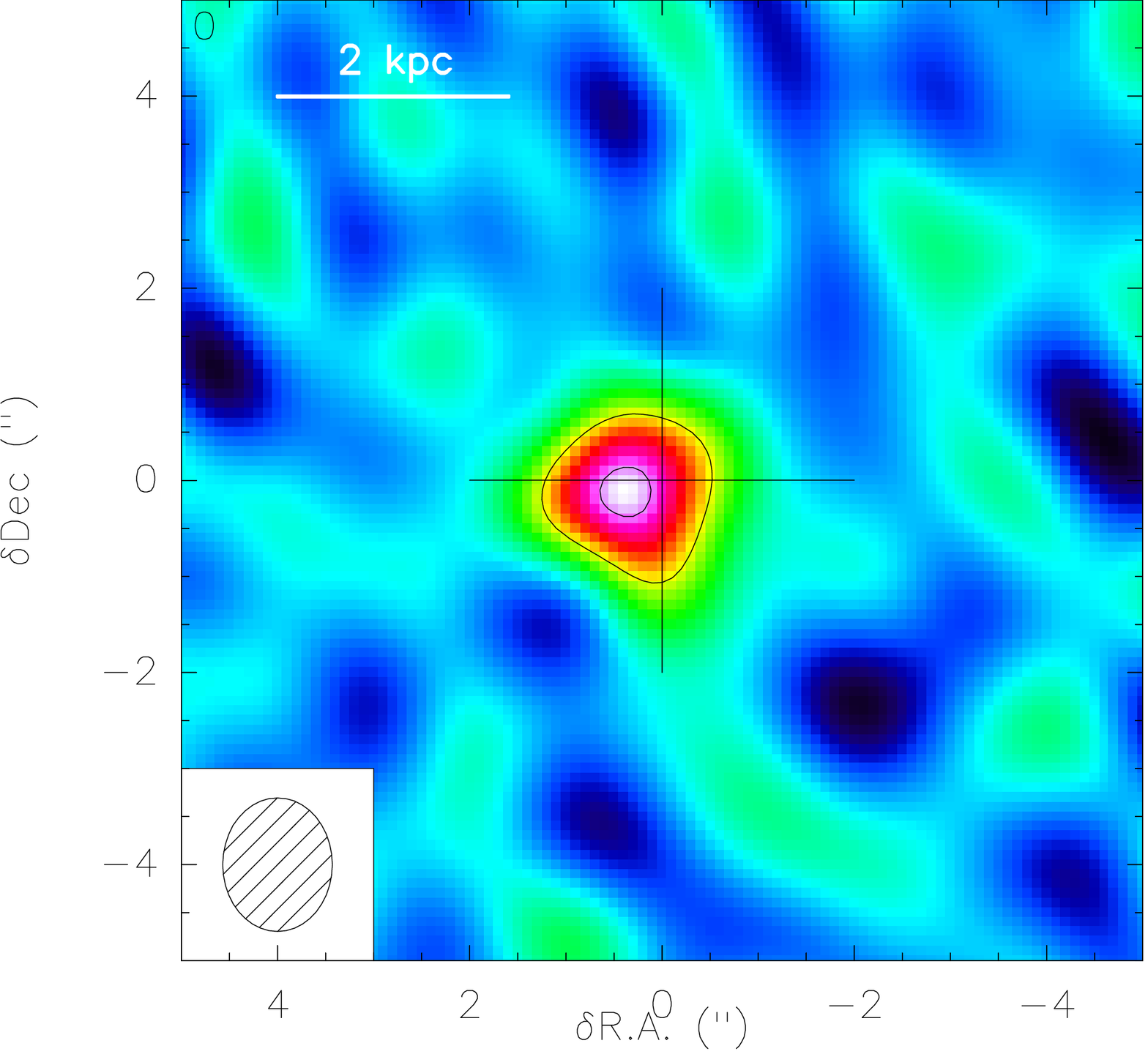}} \quad
{\includegraphics[width=.42\columnwidth, angle=270]{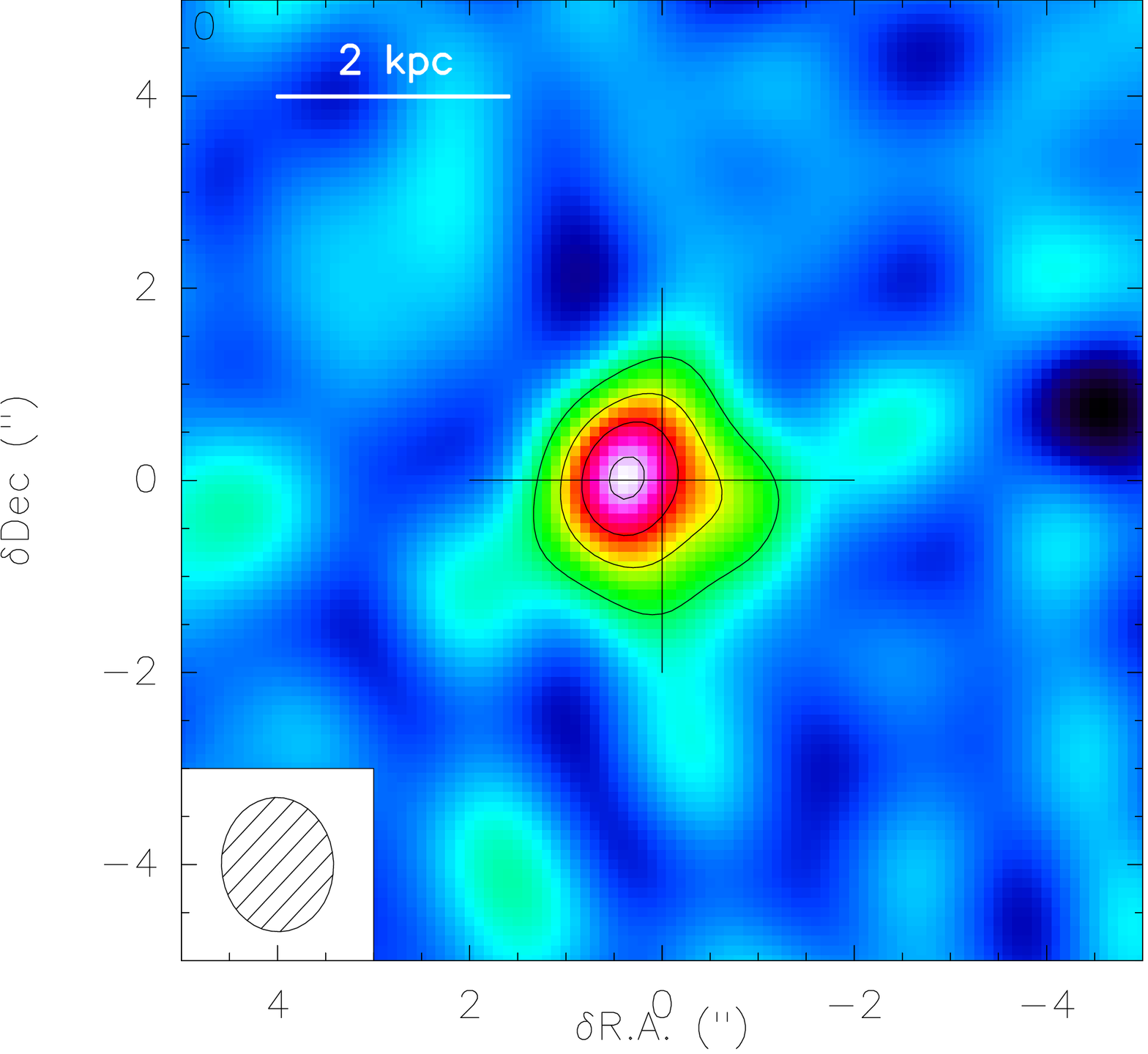}} 
\caption{IRAM PdBI maps of the CO(2-1) blue (\emph{left panel}) and red (\emph{right panel}) broad wings.
The size of the map is $10\times 10$ arcseconds. 
Contours correspond to 3$\sigma$ (1$\sigma$ = 0.6 mJy beam$^{-1}$). 
The synthesized beam size is shown at the bottom of the maps. 
%($1.6\times 1.3$ arcseconds).
The cross indicates the peak of the radio (VLBI) emission.
A tapering with uv taper = 50 m has been applied to both the maps; the
tapering cuts the visibilities with uv radius $<$ 50 m, so the maps
appear more detailed and the synthesized beam is slightly smaller than in
the original maps.
Note that the peaks 
of the wing maps appear to be slightly offset from the peak of the radio VLBI 
but are well within the beam. } \label{fig:wing21map} \end{figure}
%Figure 4: the visibility plots with Gaussian models
\begin{figure}[!]
\centering
{\includegraphics[width=.47\columnwidth]{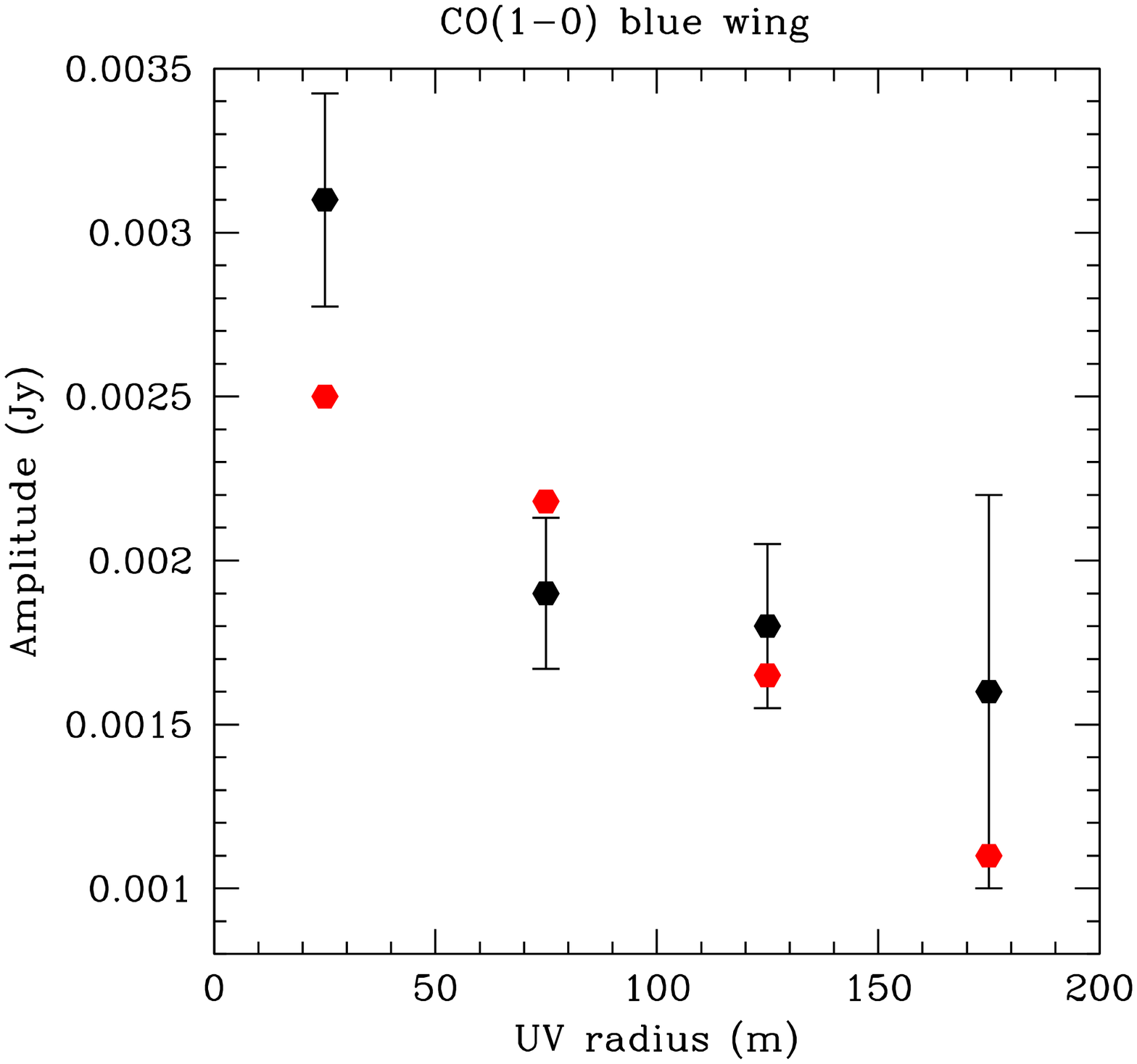}} \quad
{\includegraphics[width=.47\columnwidth]{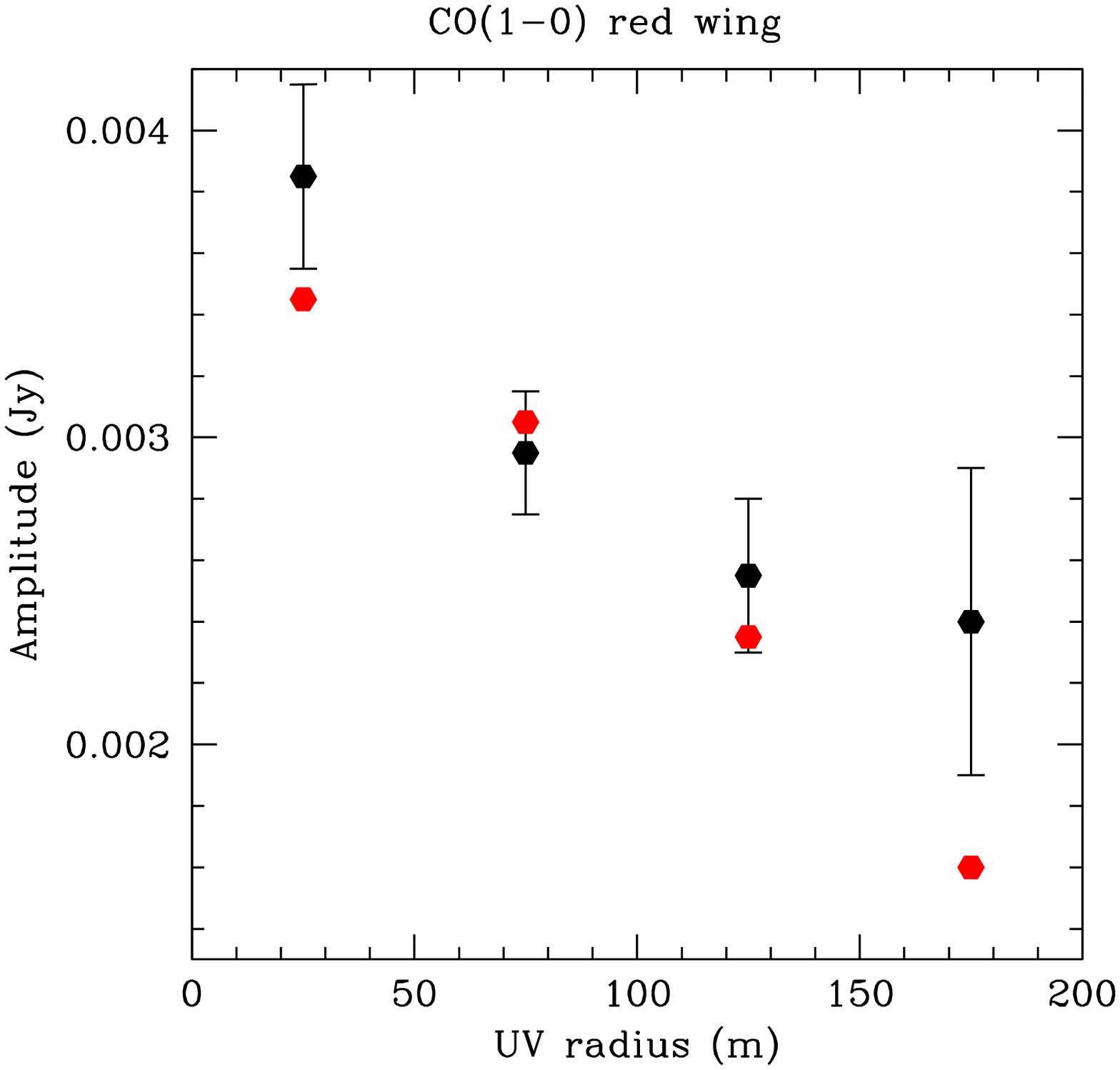}} \quad
{\includegraphics[width=.47\columnwidth]{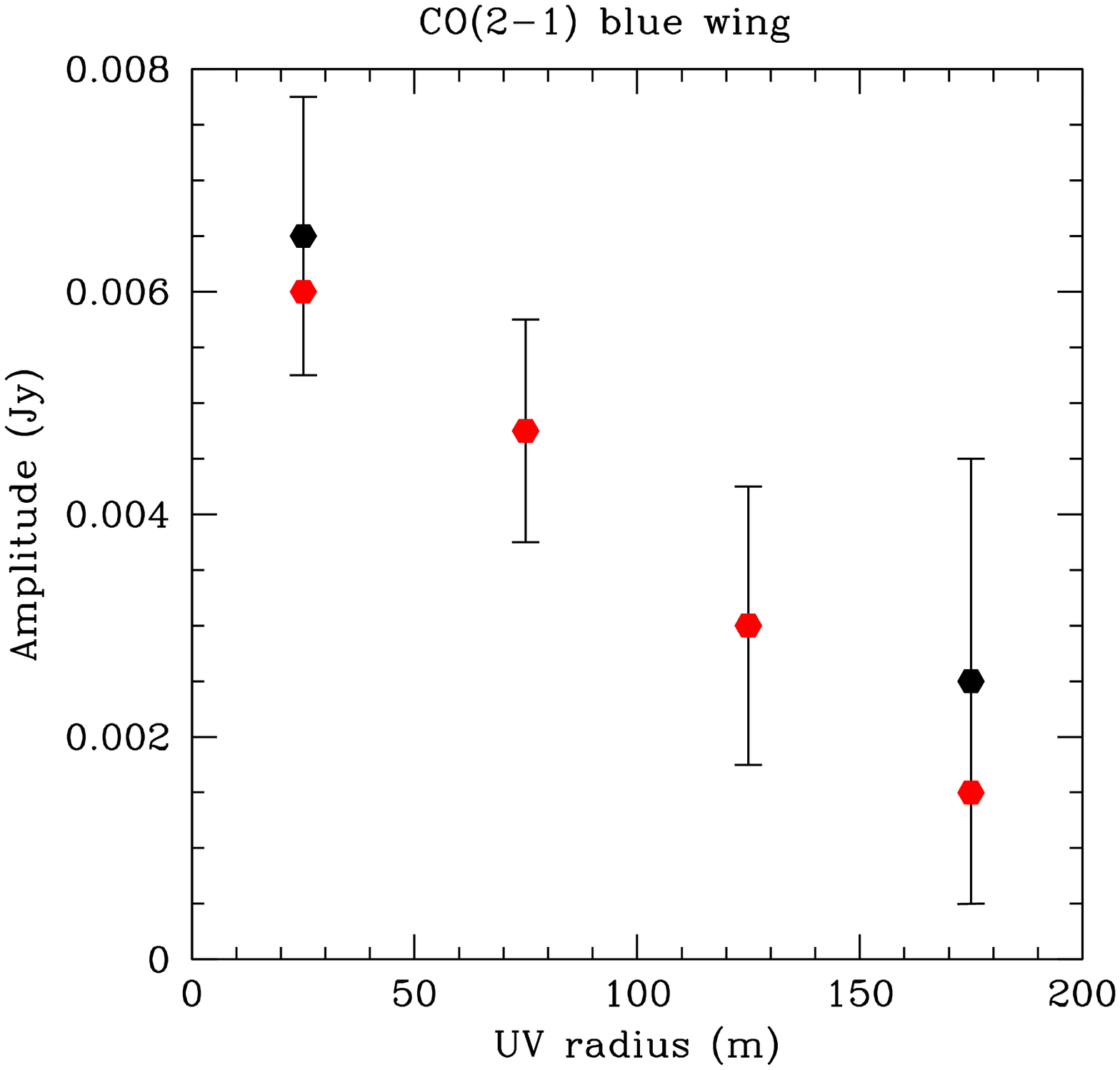}} \quad
{\includegraphics[width=.47\columnwidth]{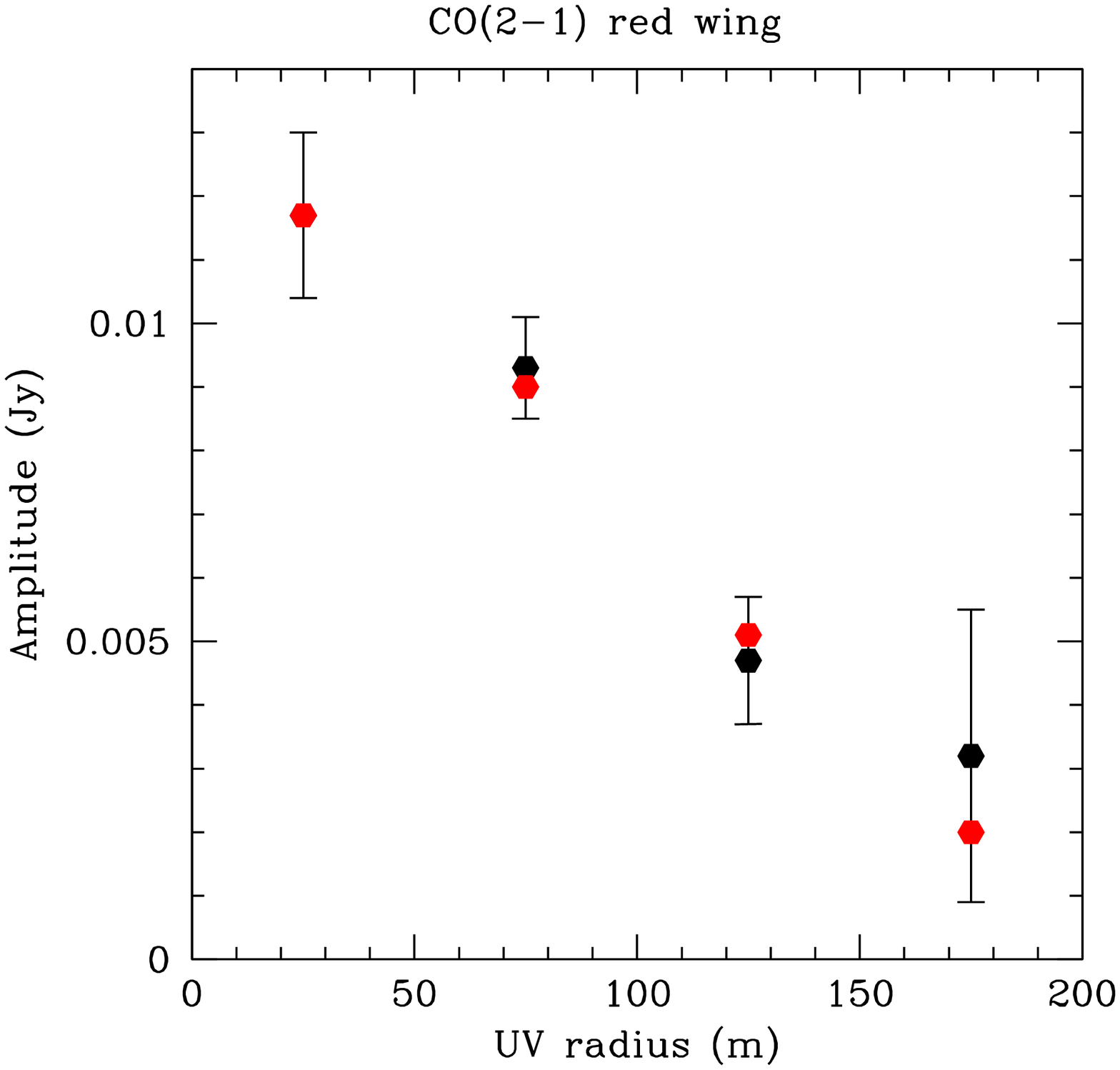}} \\
\caption{Amplitude of visibilities (Jy) plotted as a function of the uv radius for the blue 
and red wings of the CO(1-0) and CO(2-1) emission lines (black points with errorbars). 
The visibilities were binned using intervals of 50 m of uv radius, covering baselines from 10 m to 200 m.
The emission from the wings was estimated within the velocity intervals $-1000\lesssim$v(km s$^{-1}$)$\lesssim -400$
and $400\lesssim$v(km s$^{-1}$)$\lesssim 1000$.
We recall that for a point (i.e. unresolved) source the visibility amplitude is a constant function of the
uv radius. 
The red points are the results of the best fits to a circular Gaussian model.}
\label{fig:uvplots}
\end{figure}

Figures \ref{fig:wing10map} and \ref{fig:wing21map} 
show the maps of the blue and red wings of the CO(1-0) and CO(2-1) emission lines and the
half-power contours of the synthesized beams. The emission from both the broad components 
of the CO(1-0) and CO(2-1) lines is spatially resolved.
%%Fab qui diceva: ''thus improving the results by \citet{Feruglio+10}? Lo dico dopo agli uv fit
The peaks of the red and blue wings of both the CO(1-0) and the CO(2-1) emission lines are
not spatially offset from one another. This validates the hypothesis that 
these high-velocity components are due to the motion of outflowing gas, 
which moves in roughly an axisymmetric way relative to the line of sight,
instead of a rotating molecular disk.
That a large fraction of the outflow occurs along our line of sight is supported by the 
blueshifted OH absorption observed in a Herschel PACS spectrum \citep{Fischer+10}.

To quantify the extension of the wings, in Figure \ref{fig:uvplots} we plot 
the amplitude of the visibilities as a function of the uv radius 
for both the blue and red wings of the CO(1-0) and CO(2-1) emission lines.
We have binned the visibilities in baseline steps of 50 m.
The data points in Figure \ref{fig:uvplots} were fitted with both a point (unresolved) source model and a
circular Gaussian source model. The resulting reduced $\chi^2$ and the associated probability $P$ values of the
best fits are reported in Table \ref{table:uvres}, which also gives
the differences $\Delta\chi^2$=$\chi^2$(point)-$\chi^2$(Gaussian) and their associated $1-P$.
We estimated the physical extent of the outflow in terms of the FWHM 
of the best-fit circular Gaussian model fits for the uv data.
It should be noted that since we directly fitted the uv data, the fit automatically takes into account
the interferometric beam size. The resulting sizes are reported in Table \ref{table:uvres}.
We note that while the uvfit is generally good, in some cases (at some uv radii),
the Gaussian fit to the amplitudes deviates by about 1.5$\sigma$
from the observed values, this means that the spatial distribution of the flux is probably more complex
than the simple circular Gaussian approximation assumed by us.

The results in Table \ref{table:uvres} demonstrate that
both the red and blue wings of the CO(1-0) and those of the CO(2-1) lines are spatially resolved 
at high levels of significance (much higher than previously reported in Feruglio et al. 2010 for the CO(1-0)
thanks to the stronger signal on long baselines even for this transition). More specifically,
the spatial distribution of the wings is resolved with a confidence ranging from 5$\sigma$ to 11$\sigma$, depending
on the transition and the wing component.
These results prove
that the outflow extends out to the kpc scale:
FWHM$\sim$1.2 kpc in the case of the CO(1-0) transition
and FWHM$\sim$0.8 kpc for the CO(2-1) transition.
%NOTE: the fit results come from the NON BINNED UVTABLES!

Summarizing, our new data not only detect for the first time the broad wings of the CO(2-1) transition,
and resolve them spatially, 
but also allow us to estimate the extension of the CO(1-0) broad wings with much higher
accuracy than previous studies \citep{Feruglio+10}.

%Figure 4: Relative strengths of both narrow and broad CO component in Mrk 231 compared with other galaxies
\begin{figure}[!]
  \centering
  \includegraphics[width=.95\columnwidth]{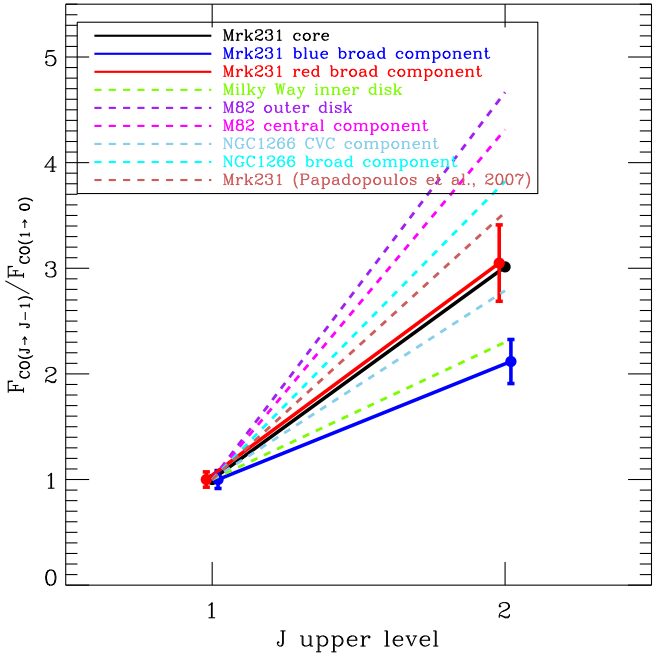}
  \caption{Relative strengths of both the narrow core (\emph{black}) and the broad wings 
(\emph{blue} and \emph{red}) of Mrk 231's CO emission lines. The ratios of the CO integrated fluxes 
were calculated using the data reported in the upper part of Table \ref{table:3fluxes}.
The CO spectral line energy distribution (SLED) of Mrk 231 obtained by \citet{Papadopoulos+07} is shown for comparison, 
along with the CO SLEDs of 
the Milky Way \citep{Fixsen+99}, M82 \citep{Weiss+05} and NGC1266 \citep{Alatalo+11}.}
\label{fig:SLED}
\end{figure}

\section{Discussion}

\subsection{The CO(2-1)/CO(1-0) outflow excitation ratio}

The physical conditions of molecular gas in galaxies, such as its kinetic temperature, volume density, and 
optical depth, can be 
derived from the CO excitation properties and, in particular, from the relative strengths of the CO emission lines.
The physical conditions of the molecular gas responsible for the \textit{narrow} components of Mrk 231's CO emission 
lines have been widely investigated. \citet{Papadopoulos+07} and \citet{Vanderwerf+10} found that the rotational 
excitation diagram of the narrow core of Mrk 231's CO lines is dominated, in the lower J levels ($J_{upper}\in[1,3]$), 
by a component of density n(H$_2)\sim 10^3$ cm$^{-3}$ and excitation temperature $T \in[55,95]$ K.
In principle, a similar analysis could be undertaken for the CO wings tracing the outflow. However, two CO transitions
are certainly not enough to investigate the excitation and physical conditions of the outflowing gas. As a consequence,
we simply compare
in the following the CO(1-0)/CO(2-1) ratio observed in the wings with the same ratio observed in the
core of the line and with other environments.

Figure \ref{fig:SLED} shows the CO(1-0)/CO(2-1) ratio for the two wings and for the line core, compared with
CO SLEDs observed in other galaxies, more specifically:
the Milky Way's inner disk $(2°.5<|l|<32°.5)$
\citep{Fixsen+99}, the outer disk and the central component of M82  \citep{Weiss+05}, both the core
and the broad component of NGC1266 \citep{Alatalo+11}, and
the CO core SLED of Mrk 231 obtained by \citet{Papadopoulos+07}.
It should be noted that the error bars in Figure \ref{fig:SLED} 
represent only the statistical uncertainties, since here we are
primarily interested in the excitation of the wings relative to
the core (hence the absolute calibration uncertainties are unimportant). However,
when comparing excitation ratios with other targets, an uncertainty
of 10\% should be added in quadrature.

The CO(1-0)/CO(2-1) ratios observed in the red wing and in the narrow core
are consistent with each other. The blue wing is less 
excited than the core of the line.
Taken at face value, this result does not support AGN feedback models where the outflow is
generated by a radiation-driven shock (``blast wave models''),
which, besides accelerating the ISM, should greatly increase its temperature
and density \citep[e.g. ][]{Lapi+05,Zubovas+12}. However, it is possible that a high excitation component
of the outflow may be revealed by high-J transitions. Therefore, sensitive and wide band
observations of higher CO transitions are required to further investigate this issue.

However, it should be noted that, since the CO(2-1) wings are
more compact than the CO(1-0) wings, the CO(2-1)/CO(1-0) ratio is likely
to vary radially. More specifically, the CO(2-1)/CO(1-0) ratio 
should be higher
in the central region than the outer region.
Higher angular resolution maps would be required to
properly map the CO(2-1)/CO(1-0) ratio and disentangle the beam-smearing effects.
However, on the basis of the circular Gaussian modeling of the spatial
distribution of the two wings in the two transitions, we estimated that
within the central 0.7~kpc (FWHM of the blue wing) the CO(2-1)/CO(1-0) ratio
of the blue wing increases to 3.5$\pm$0.8, which is however still 
consistent with the excitation observed in the core of the lines
($\rm F_{CO(2-1)}/F_{CO(1-0)}=2.9$). The CO(2-1)/CO(1-0)
flux ratio of the red wing increases
in the same inner region to 4.0$\pm$0.8, which starts to be inconsistent with
the excitation in the core of the line, although marginally. On the other
hand, in the region outside the central 0.7~kpc the CO(2-1)/CO(1-0) ratio
drops significantly, below 2, well below the excitation observed in the core of
the line. Summarizing, the outflowing gas associated with the blue wing does 
not show any evidence of higher excitation relative to the core of the line,
both globally and even in the central region, indicating that this outflowing
gas is unaffected by shocks. Within the central region, the gas associated
with the red wing does show marginally higher excitation, suggesting
that this component of the outflowing gas may be affected by some shocks.

The high Mach number inferred by \cite{Feruglio+10}, based on the outflow velocity of the molecular clouds,
is also indicative of shocks. However, we also note that high outflow velocities
do not necessarily translate into a large Mach number, since the medium in which the clouds are embedded may be
outflowing with the same velocity (as appears to be the case from the similar velocity of the ionized gas found
by Rupke \& Veilleux 2011), hence resulting in a smaller differential velocity between the molecular
clouds and the surrounding gas.

Another important implication of our result is that the CO-to-H$_2$ conversion factor in the wings may not
differ too much from the bulk of the gas in the host galaxy traced by the narrow CO component
(i.e. a ULIRG-like conversion factor). In \citet{Feruglio+10}, we had been conservative by assuming, for the
outflow, a conversion factor $\rm \alpha = 0.5~M_{\odot}~(K~km~s^{-1}pc^2)^{-1}$ (i.e. one-tenth of the Galactic value and
40\% lower than the value assumed for ULIRGs);
this is the lowest conversion factor found in different locations of M82, including its molecular outflow
\citep{Weiss+01}. The finding that the CO excitation in the outflow does not differ significantly from (or is even lower)
than
the core of the line, tracing the bulk of the gas in this ULIRG, suggests that the conversion factor
is also similar (i.e. $\rm \alpha = 0.8~M_{\odot}~(K~km~s^{-1}pc^2)^{-1}$). This implies that the outflow
molecular mass and outflow rate given in \citet{Feruglio+10} are underestimated,
and are probably higher by a factor of about 1.4.

We note that in the jet-driven outflow observed in NGC~1266 by \citet{Alatalo+11} the excitation of the CO broad
components tends to be higher than in the central velocity component (as illustrated in Fig. \ref{fig:SLED}),
although not strongly. This is not unexpected. In this case, the outflow
 is totally driven by jet-induced shocks, which unavoidably heat the ISM and therefore produce higher
CO excitation. The outflow CO excitation differences between Mrk~231 and NGC~1266 further highlight that
the driving mechanism in the two sources are different, although both associated with an AGN.

Finally, we note that the Mrk~231 SLED of the core obtained by \citet{Papadopoulos+07} by means of single dish
observations is slightly steeper than the one obtained by us. We ascribe the difference partly to calibration
uncertainties and partly to uncertainties in subtracting the continuum and broad wings with previous
narrow band spectra (especially in single dish data where the baseline instabilities may be problematic).

%Table 5: UV fit results with circular gaussian function model
%%% ATTENZIONE: conversione=0.828 kpc/arcsec
\begin{table*}[!]
\caption{Best-fit results of the visibility versus uv radius plots.}
\label{table:uvres}
\begin{center}
\begin{tabular}{lcccccc}
\hline\midrule
 Line 	& Velocity range    & Point		& Gaussian 			& $\Delta\chi^{2\dag}$ 				& FWHM         & FWHM $^\ddag$     \\
		&  (km s$^{-1}$)   & source model    	& source model        	 &                                   		& (arcsec) 	& (kpc)  \\
\midrule
 CO(1-0)   	 & [$-1000, -400$]  & $\chi^2_r$ = 4.14 & $\chi^2_r$ = 1.98 & $\Delta\chi^2$ = 6.48 			 &  $(1.52 \pm 0.31)$  & $(1.26 \pm 0.25)$  	\\
 blue wing       &                      & $P$ = 6.1E-03     & $P$ = 0.11        & $(1-P)$ = 0.99				&			&			\\
\midrule
 CO(1-0)    	 & [$400, 1000$]    & $\chi^2_r$ = 4.26 & $\chi^2_r$ = 1.74 & $\Delta\chi^2$ = 7.54	 			& $(1.37 \pm 0.22) $  & $(1.13 \pm 0.18)$      \\
 red wing        &			& $P$ = 5.2E-03 &  $P$ = 0.16 		& $(1-P)$ = 0.99				&		     &		    \\
\midrule 
CO(2-1)   	& [$-1000, -400$]  & $\chi^2_r$ = 1.69 & $\chi^2_r$ = 0.14 & $\Delta\chi^2$ = 4.65 			& $(0.80 \pm 0.16)$ & $(0.66 \pm 0.14)$ \\	
 blue wing	&			& $P$ = 0.17 	& $P$ = 0.94 		& $(1-P)$ = 0.97			&  		 &		 	\\
\midrule
CO(2-1)   	& [$400, 1000$] 	& $\chi^2_r$ = 8.94 &  $\chi^2_r$ = 0.19  & $\Delta\chi^2$ = 26.24			&   $(1.01 \pm 0.09)$ & $(0.84 \pm 0.07)$   \\
 red wing	&			& $P$ = 6.5E-06 	& $P$ = 0.90	 & $(1-P)\simeq$ 1.00 				&			&		\\
\midrule
\end{tabular}
\end{center}

\textbf{Notes:} 
$^\dag$  $\Delta\chi^2$=$\chi^2$(point)-$\chi^2$(Gaussian)
$^\ddag$ The adopted cosmology and redshift result in a spatial scale of $0.828$ kpc arcsec$^{-1}$.
\end{table*}

%Figure 5: Plot of size vs critical density
\begin{figure}[!]
  \centering
  \includegraphics[width=.9\columnwidth]{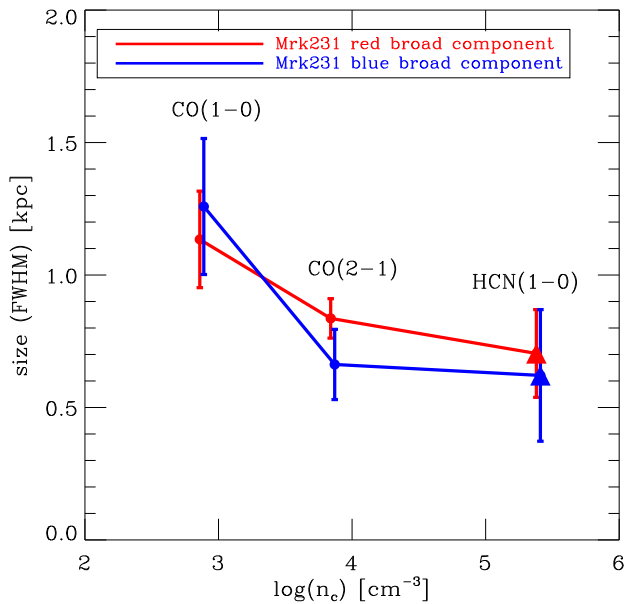}
  \caption{Size (FWHM) of the outflow plotted as a function of the critical density of the
corresponding molecular transition used as a wind tracer. Our CO(1-0) and CO(2-1) IRAM PdBI 
data are represented by filled circles, while the HCN(1-0) observations by \citet{Aalto+11} are denoted by filled
triangles. Critical densities were calculated for a molecular medium at $T = 100$ K. 
Note that the same conversion factor of 0.828 kpc arcsec$^{-1}$ has been applied to all data in order
to allow comparisons. }
\label{fig:extension}
\end{figure}

\subsection{Extension of the outflow}

As mentioned above, our new data have allowed us to constrain the
size of the molecular outflow traced by the CO transitions with
unprecedented accuracy, enabling analyses that have not been possible
until now. The results presented in the previous section clearly
reveal that the extension of the outflow traced by the CO(1-0) transition
is significantly larger
than observed in the CO(2-1) and HCN(1-0)
transitions. Fig.~\ref{fig:extension} shows the extension of the outflow
as a function of the critical density of the transition used to trace the
outflow, where the red and blue symbols indicate the trends relative
to the red and blue wings, respectively. There is a clear trend for the size
of the outflow to decrease as a function of the critical density, at least in the range
$\rm 10^3 < n_{cr} <10^4 ~cm^{-3}$ (while at higher densities it is more difficult
to assess this trend owing to the large errorbars in the size of the HCN wings).

A radial decreasing density of the outflowing molecular gas is also
consistent with the Herschel measurements by Fischer et al. (2010) and Sturm et al. (2011),
and in particular with the blueshifted absorption profile of the far-IR molecular
transitions observed in the PACS spectra (OH 79$\mu$m, OH 119$\mu$m, $^{18}$OH 120$\mu$,
H$_2$O 79$\mu$m), as discussed in those papers.

This result is consistent with the
prediction of \citet{Narayanan+08}, who studied the response of CO morphology
to starburst and AGN feedback-driven winds in galaxy mergers.
According to them, the emission from higher CO transitions (e.g. CO(2-1))
broadly follows the distribution of the CO(1-0) emission, but 
remains slightly more compact. This is because, according to their
model, denser clouds (preferentially
traced by higher transitions) have shorter lifetimes along the outflow and tend to
evaporate into a more diffuse molecular phase, traced by the CO(1-0) transition.

The result could also possibly support the scenario proposed by Hopkins \& Elvis (2010), where
dense clouds invested by an even mild AGN-driven wind develop Kelvin-Helmholtz instabilities
causing them to expand and accelerate more easily by means of radiation pressure. In the latter scenario,
one would expect diffuse clouds to be driven to larger distances than the denser
clouds. However, Kelvin-Helmholtz instabilities should also develop shocks (although
much milder than those expected in ``blast wave models'') that should show up in higher gas
excitation, which is not the case for the blue wing, but may be for the red wing.

Simple acceleration by radiation pressure on the dust in molecular clouds
\citep{Fabian+09}, without invoking any scenario associated with shocks, may 
not only explain the modest CO excitation in the wings, but also the size difference
between the CO(1-0) and CO(2-1)+HCN emission components. More diffuse clouds
(preferentially traced by CO(1-0)) are indeed characterized by a higher ratio between the outward radiation
pressure and inward gravitational pull than dense clouds, which are preferentially traced by CO(2-1)+HCN
emission. More specifically, for clouds whose dust is optically thick to UV/optical
radiation, the ratio $\rm F_{rad}/F_{grav}$ is inversely proportional to the column density of the
cloud $\rm N_H$, which is higher in dense clouds.
As a consequence, diffuse clouds can be expelled to larger distances.
Radiation pressure on the dust in molecular clouds can in principle also develop shocks,
but this depends on whether the medium in which the molecular clouds are embedded is
comoving with them.

We have discussed so far only the global size of the emitting regions of
the two CO transitions obtained by fitting the uv data with a simple circular
Gaussian. At our resolution, the maps do not show much structure. The
only relevant structure at the 3$\sigma$ level is possibly an extension of the CO(1-0)
blue wing (Fig.2, left) a few arcseconds to the north. This could imply that there is an
association between some of the molecular outflow (a minor part of it) and the
northern region, where the radio jet affects the neutral gas outflow as
inferred by \citet{Rupke+11} from the blue-shifted Na I D kinematics map.
\citet{Aalto+11} detected a plum extending to the north in the HCN(1-0)
red-shifted wing map and also inferred a link between this feature and the
jet-influenced neutral wind detected by \citet{Rupke+11}. However, we propose that
this association is incorrect, since the jet-accelerated gas is detected
in terms of {\it blueshifted} Na I D absorption, so any association should be seen
in the molecular {\it blue} wings, not the red wings. The possible
interaction between part of the molecular outflow and the radio-jet should
however be investigated using higher-resolution data.

Finally, we note that the size ($\sim$2 arcsec) of the (blueshifted) ``nuclear
wind'', which is the region identified by Rupke \& Veilleux (2011) where the
outflow is dominated by a radiation-pressure drive wind, is close to the size
($\rm 1.5\pm 0.3$ arcsec) of the molecular outflow determined by us from the blue
wing of the CO(1-0) component.

\section{Conclusions}

We have presented new broad-band IRAM PdBI observations of the CO(2-1) transition 
in Mrk~231, which is the ULIRG hosting the closest quasar known. For the first time,
our observations reveal the broad wings of the (2-1) transition extending to velocities of
FWZI$\sim$1500~km s$^{-1}$, which trace the same molecular wind revealed by previous CO(1-0)
PdBI observations and PACS far-IR spectroscopy of OH transitions. The CO(2-1) wings
are spatially resolved, with an extension of about 0.8~kpc.

We have also obtained new CO(1-0) observations, by significantly improving the signal at high angular resolution
relative to previous works, which allows us to determine more accurately the
size of the molecular outflow traced by this transition. The CO(1-0) broad wings
are spatially resolved with a significance that is much higher (5--7$\sigma$) than in previous
observations. We measure an extension of the CO(1-0) wings of about 1.2~kpc, which is significantly
larger than observed for the CO(2-1) wings.
The extension of the CO(1-0) wings is also significantly larger than the extension
of the HCN(1-0) wings, tracing high density gas in the outflow detected by previous
observations. More specifically, we show that the size of the outflow anticorrelates
with the critical density of the transitions used to trace the outflow.

These results are consistent with the scenario where denser clouds (traced by the
CO(2-1) and HCN(1-0) broad wings) have shorter lifetimes
along the outflow and evaporate into a more diffuse molecular component (traced by CO(1-0)) at larger
radii, as predicted by some feedback models.

We also find that the CO excitation in the outflow, as traced by the CO(2-1)/CO(1-0)
ratio of the broad wings, does not differ significantly from that of the gas in the
bulk of the galaxy, as traced by the core of the CO lines. In the blue wing, the
excitation is lower than in the core of the line. Taken at face value, this result
in inconsistent with models where the molecular outflow is driven by a shock
wave generated by the interaction of a radiation pressure-driven nuclear wind with the
ISM of the host galaxy.
We favor a scenario where gas clouds are directly accelerated by the radiation pressure
on dust. This scenario does not need to invoke shocks, hence the CO excitation is 
unaffected. Moreover, in this scenario low density clouds (traced by CO(1-0)) are accelerated
more efficiently, hence reaching larger distances than dense clouds (traced
by CO(2-1) and HCN).
However, we also note that in the inner region ($\rm R<0.3\,kpc$), the CO(2-1)/CO(1-0) ratio is indeed
slightly higher, which is indicative of some shock contribution.

We marginally detect an extension of the CO(1-0) blueshifted wing to the
north, where previous studies have
found evidence that the atomic neutral outflow, traced by blueshifted Na I D
absorption, is influenced by the radio jet.
This suggests that the radio jet may also contribute to the acceleration of
part of the molecular outflow,
although far less significant than the global molecular outflow.

\begin{acknowledgements}
We thank Andrea Lapi for helpful suggestions and discussions. We are grateful to
the IRAM staff in Grenoble for helping with data reduction and calibration.\\
This work is based on observations carried out with the IRAM Plateau de Bure Interferometer. 
IRAM is supported by INSU/CNRS (France), MPG (Germany), and IGN (Spain).
This work was supported by ASI/INAF contract I/009/10/0.

\end{acknowledgements}

\bibliography{ref}
\bibliographystyle{aa}

\end{document}